\documentclass{PoS}
\usepackage[pdftex]{graphicx}
\usepackage{subfigure}
\usepackage{wrapfig}

\title{The IceCube Upgrade - 
	Design and Science Goals}

\ShortTitle{IceCube Upgrade}
\author{
The IceCube Collaboration\footnote{For collaboration list, see PoS(ICRC2019) 1177.}\\
{\itshape \href{http://icecube.wisc.edu/collaboration/authors/icrc19_icecube}{http://icecube.wisc.edu/collaboration/authors/icrc19\_icecube}}\\
E-mail: \email{aya@hepburn.s.chiba-u.ac.jp}
}
\abstract{
The IceCube Neutrino Observatory at the geographic South Pole has reached a number of milestones in the field of neutrino astrophysics. The achievements of IceCube include the discovery of a high-energy astrophysical neutrino flux, and the temporal and directional correlation of neutrinos with a flaring blazar. The IceCube Upgrade, which will be constructed in the 2022/23 Antarctic Summer season, is the next stage of the IceCube project. The IceCube Upgrade consists of seven new columns of photosensors, densely embedded near the bottom center of the existing cubic-kilometer-scale IceCube Neutrino Observatory.  An improved atmospheric neutrino event selection efficiency and reconstruction at a few GeV can be achieved with the dense infill of the Upgrade's photosensor array. The Upgrade will provide world-leading sensitivity to neutrino oscillations and will enable IceCube to take unique measurements of tau neutrino appearance with a high precision. Furthermore, the new array will also improve the existing IceCube detector. The Upgrade strings will include new calibration devices designed to deepen the knowledge of the optical properties of glacial ice and the detector response. The improved calibration resulting from the Upgrade will be applied to the entire archive of IceCube data collected over the last 10 years, improving the angular and spatial resolution of the detected astrophysical neutrino events. 	Finally, the Upgrade represents the first stage in the development of IceCube-Gen2, the next-generation neutrino telescope at the South Pole.

\vspace{4mm}
{\bfseries Corresponding author:}
\speaker{Aya Ishihara}$^{\;1}$\\
{$^{1}$ \itshape Dept. of Physics and Institute for Global Prominent Research, Chiba University, Chiba 263-8522, Japan}

}

\FullConference{36th International Cosmic Ray Conference -ICRC2019-\\
		July 24th - August 1st, 2019\\
		Madison, WI, U.S.A.}

\begin{document}

\section{What's the IceCube Upgrade?}\label{sec:introduction}
\begin{figure}[b]
	\begin{center}	
		\vspace{0.4cm}
		\includegraphics[width=1.1\linewidth]{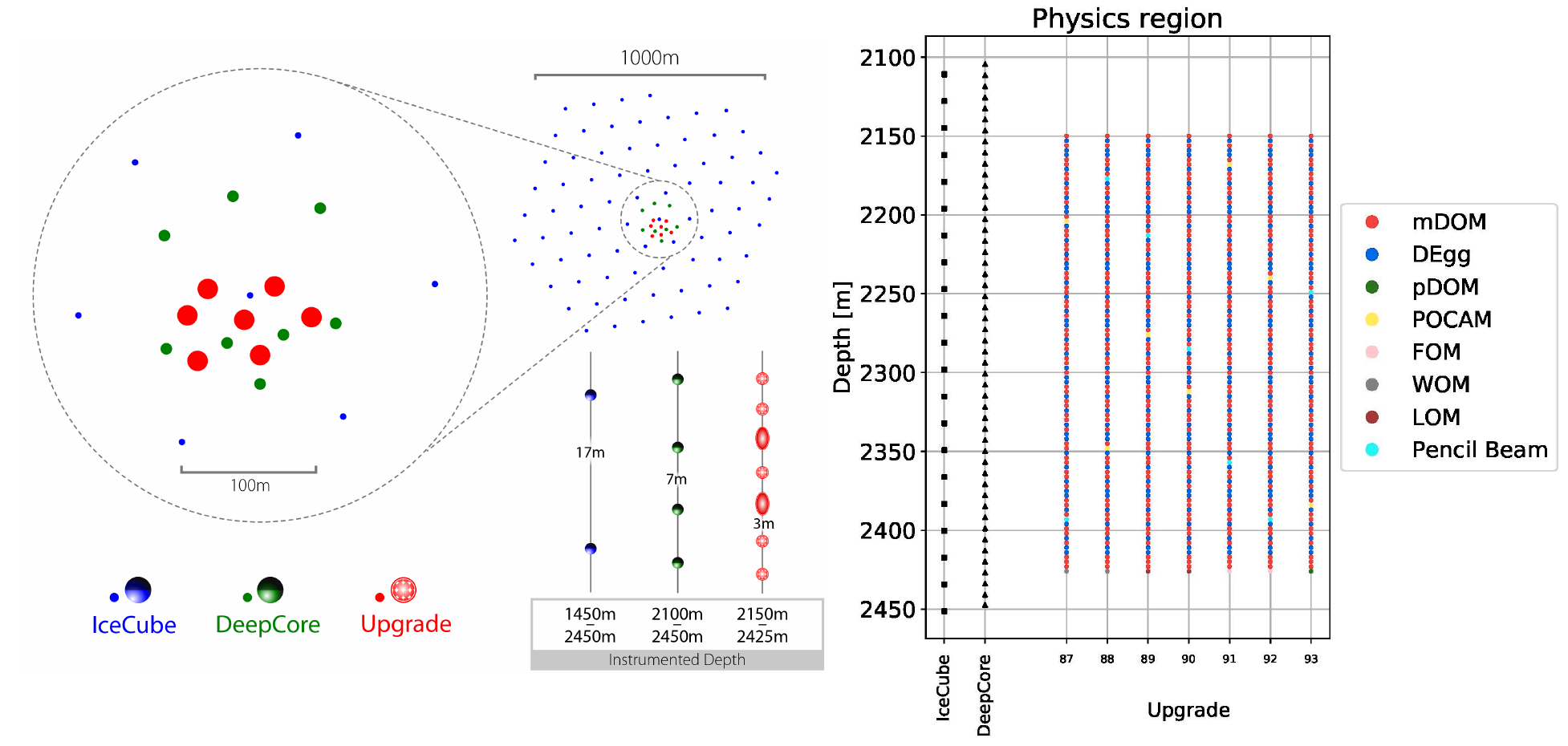}	
		\caption{The Upgrade array geometry. Red marks on the left panel shows the layout of the 7 IceCube Upgrade strings with the IceCube high-energy array and its sub-array DeepCore. The right panel shows the depth of sensors/devices for the IceCube Upgrade array (physics region). The different colors represent different optical modules and calibration devices. The Upgrade array extends to shallower and deeper ice regions filled with veto sensors and calibration devices (special calibration regions).}
	\end{center}
	\label{fig:design}
\end{figure}
The IceCube Neutrino Observatory was completed at the South Pole in 2011. IceCube has led to many new findings in high-energy astrophysics, including the discovery of an astrophysical neutrino flux and the temporal and directional correlation of neutrinos with a flaring blazar~\cite{IC_Summary}. It has defined a number of upper-limits on various models of the sources of ultra-high energy cosmic rays, as well as measurements on the fundamental high-energy particle interactions, such as neutrino cross sections in the TeV region~\cite{IC_Summary2}. 

IceCube uses glacial ice as a Cherenkov medium for the detection of secondary charged particles produced by neutrino interactions with the Earth. The distribution of Cherenkov light measured with a 1~km$^3$ array of 5160 optical sensors determines the energy, direction, and flavor of incoming neutrinos. Although the South Pole is considered one of the world's most harsh environments, the glacial ice $\sim$2 km below the surface is a dark and solid environment with stable temperature/pressure profiles ideal for noise sensitive optical sensors. IceCube has recorded detector uptime of more than 98\% in the last several years. While it has been 15 years since the first installation of the sensors, an extremely low failure rate of the optical modules has also been observed, demonstrating that the South Pole is a suitable location for neutrino observations.

The IceCube Upgrade will consist of seven new columns of approximately 700 optical sensors, called strings, embedded near the bottom center of the existing IceCube Neutrino Observatory. As illustrated in Fig.~\ref{fig:design}, the "Upgrade" consists of a 20~m (horizontal) $\times$ 3~m (vertical) grid of photon sensors at the depths between 2150~m and 2425~m, referred to as the physics region, where the glacial ice is the clearest and the atmospheric muon background is low. The optical modules to be installed in the physics region are the Multi-PMT Digital Optical Module (mDOM)~\cite{MDOM} and the Dual optical sensors
in an Ellipsoid Glass for Gen2 (D-Egg)~\cite{DEGG}, both designed to improve the photon detection efficiencies and the calibration capability of the detector. %
The deployment of the Upgrade array is planned for the 2022/23 Antarctic Summer season with an additional cabling work for the surface DAQ system planned for 2021/2022.

The Upgrade also provides new abilities to calibrate the existing detector for reanalyses of archival data, as well as a development platform for the future IceCube-Gen2; the 8 km$^3$ scale next-generation neutrino detector.
 Each of the optical modules encloses several calibration devices, such as fast ($\sim$5~ns FWHM) LEDs, CCD cameras \cite{CAMERA}, on-board pressure, temperature, magnetic field sensors, and accelerometers. In addition to the optical modules, a significant number of light-emitting stand-alone devices for the calibration of the glacial ice and the {\it in situ} detector responses, namely the POCAM~\cite{POCAM} and the Pencil Beam, and acoustic emitters~\cite{acoustic} are embedded in the Upgrade array. 
Below and above the physics region, there are sparsely (vertical distance of $\sim$25~m) instrumented regions for the purpose of the atmospheric muon veto, calibration, and "research and development" (R\&D) of the sensors specifically designed for the IceCube-Gen2 (special calibration regions). In the special calibration regions, mDOMs and D-Eggs as well as PDOMs~\cite{PDOM}, which are refurbished sensors using the same glass sphere and PMTs with newly developed electronics, will be installed for a better understanding of {\it in situ} responses of the current IceCube optical sensors. While the drilling procedure is updated from what was used for the IceCube, one of the new holes will be kept as similar as possible to the original IceCube holes, so that our understanding of optical properties in the old holes are improved. In addition, there will be performance studies of the R\&D sensors \cite{WOM_LOM_FOM}. These new types of optical sensors, developed to fit into possible narrow holes in Gen2 with a diameter of $\sim$15~cm (current hole diameter is $\sim$50~cm) to reduce construction costs, will be studied. 

\section{Science capability of the Upgrade array}\label{sec:science}
The primary reason of the exceptional science potential of IceCube, DeepCore, and Upgrade is their significant size and the cold and deep environment in which they operate, which provide extremely low rates of cosmic ray muons and detector noise. As a result of using a naturally existing Cherenkov medium, the Upgrade array, which makes up only a small section of the whole IceCube Observatory, consists of 2~MT of newly instrumented glacial ice. Such a large mass significantly reduces the statistical uncertainties in neutrino oscillation analyses. In the Upgrade, the science capabilities are further extended by the increased number of photons measured with the dense high-efficiency photosensors. 
These can record a larger number of photons from secondary charged particles induced in the $O(1\sim10$ GeV) neutrino interactions, leading to a more accurate event reconstruction in an energy region lower than main energy region of DeepCore. The Upgrade enables the detection of tau neutrino appearances with high precision. It will be the world's most stringent test on the unitarity of the Pontecorvo-Maki-Nakagawa-Sakata (PMNS) matrix, which is useful because any deviations from unitarity provide evidence for new physics beyond the Standard Model. 
Not only will neutrino oscillation results be improved by the Upgrade, but the newly calibrated data set immediately enhances IceCube's sensitivity to high-energy cosmic neutrino fluxes. The improved calibration constants obtained with the Upgrade will be applied to the entire IceCube data archive collected over 10 years. Furthermore, it has been demonstrated that the Upgrade array will enhance the sensitivity to dark matter~\cite{wimp}, which could not be discussed in this article due to space constraints. 

\subsection{Neutrino oscillations}\label{subsec:neutrino}
\begin{figure}[t]
	\begin{minipage}{\textwidth}
		\centering \hspace{3cm}
		\subfigure{\includegraphics[width=1.08\linewidth]{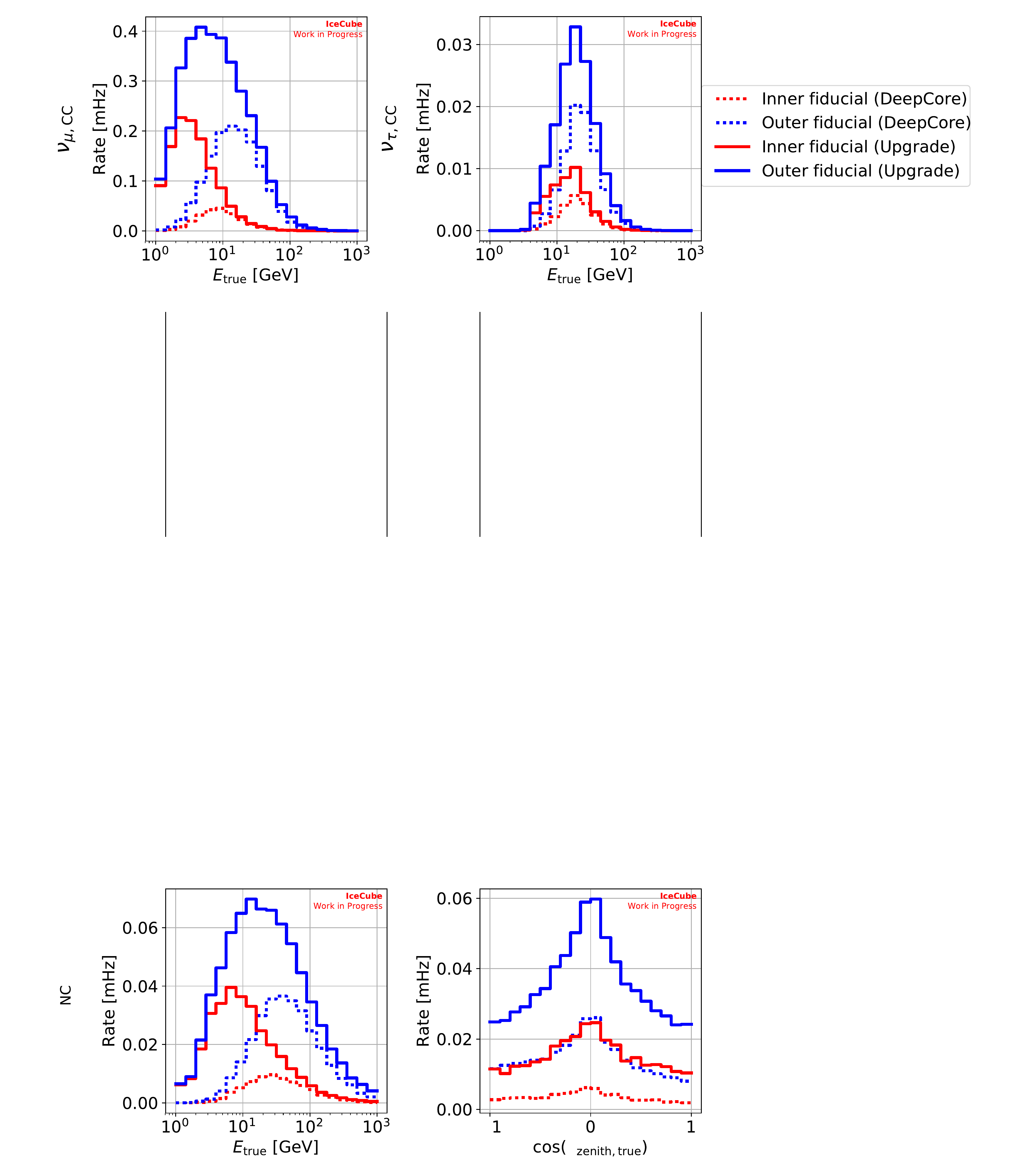}\label{fig:fig1}}
	\end{minipage}\\
	\hfill
	\begin{minipage}{\textwidth}
		\subfigure{\includegraphics[width=0.9\linewidth]{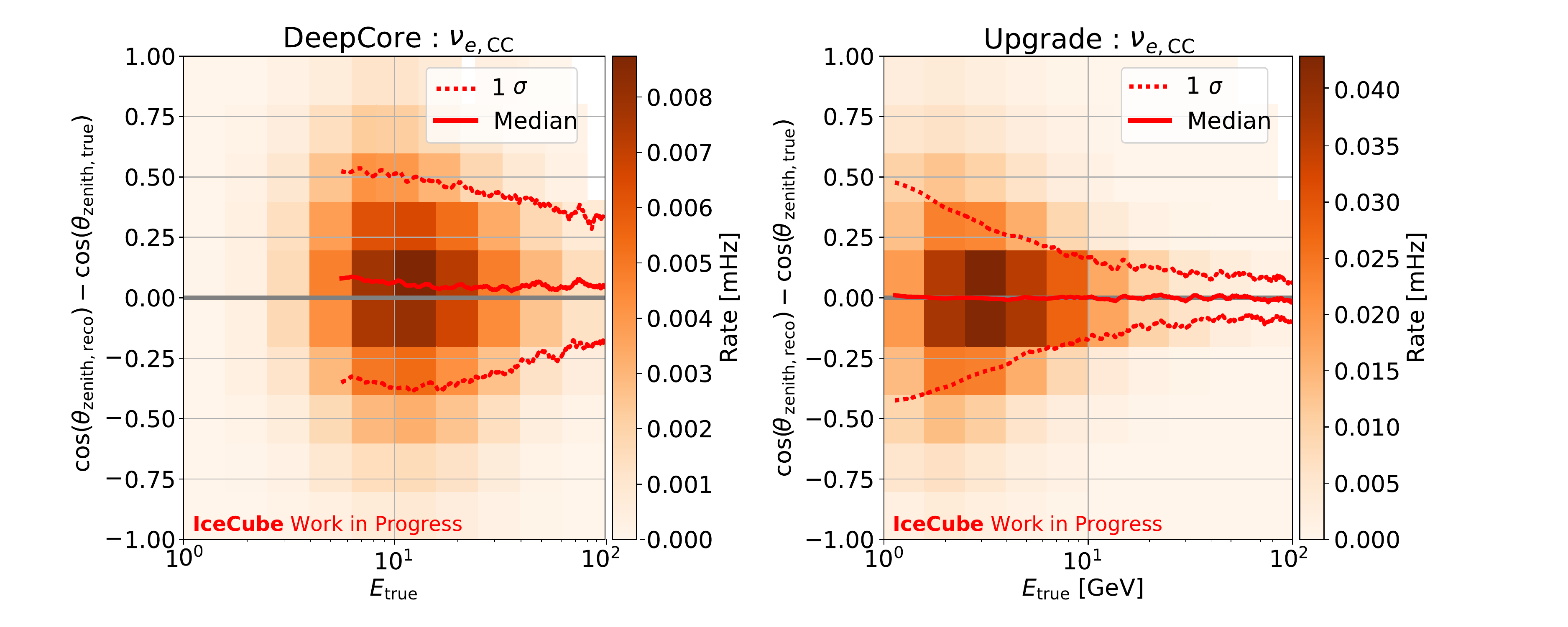}\label{fig:fig3}}
	\end{minipage}\hfill
	\caption{(Upper) Fully contained atmospheric $\nu_\mu$ and $\nu_\tau$ rates in the DeepCore and Upgrade arrays via charged current (CC) interactions at analysis level event selection. Standard oscillations are included. Dotted lines are events in DeepCore and solid lines are a new selection of events in the Upgrade. The "inner fiducial" is defined as a cylindrical volume with a 50~m radius and a height of 275~m and the "outer fiducial" is a similar volume of radius 145~m, both centered in the DeepCore and Upgrade array. Red and blue lines represent neutrino events with their vertices within the inner and outer fiducial volumes, respectively. The current sensitivity studies are performed using samples in the inner fiducial only.
		(Lower) Cascade event cos(zenith angle) reconstruction resolution for CC $\nu_e$ interactions. The reconstruction in the Upgrade array (lower right) is shown as well as that in DeepCore  (lower left). Colors indicate the distribution of reconstructed events. A factor of 3 or better improvement in zenith angle reconstruction compared with DeepCore sample are expected in the energy range relevant for the neutrino oscillation analysis.}
		\label{fig:selection}
\end{figure}
\begin{figure}[t]
	\begin{minipage}{0.48\textwidth}
	\centering 
	\subfigure{\includegraphics[width=\linewidth]{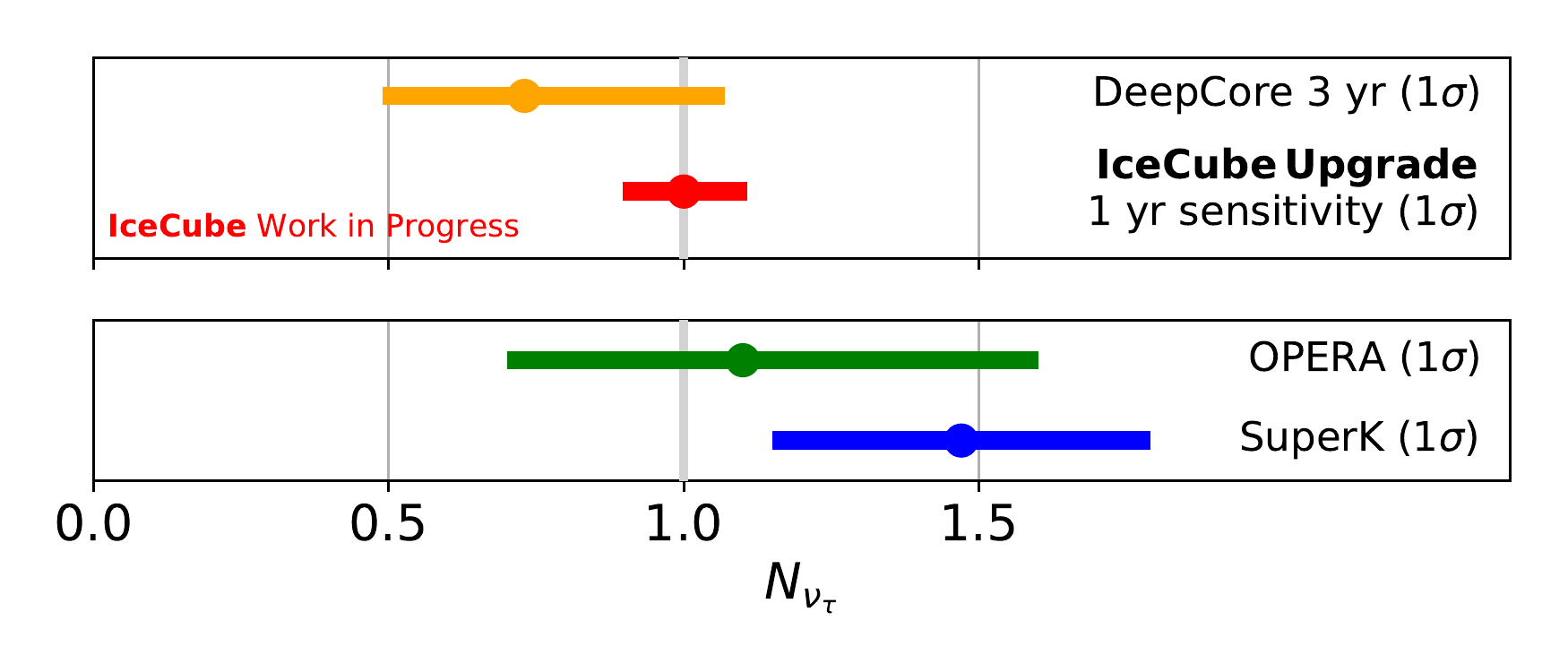}\label{fig:1}}
	\end{minipage}
	\begin{minipage}{0.53\textwidth}
	\centering
	\subfigure{\includegraphics[width=\linewidth]{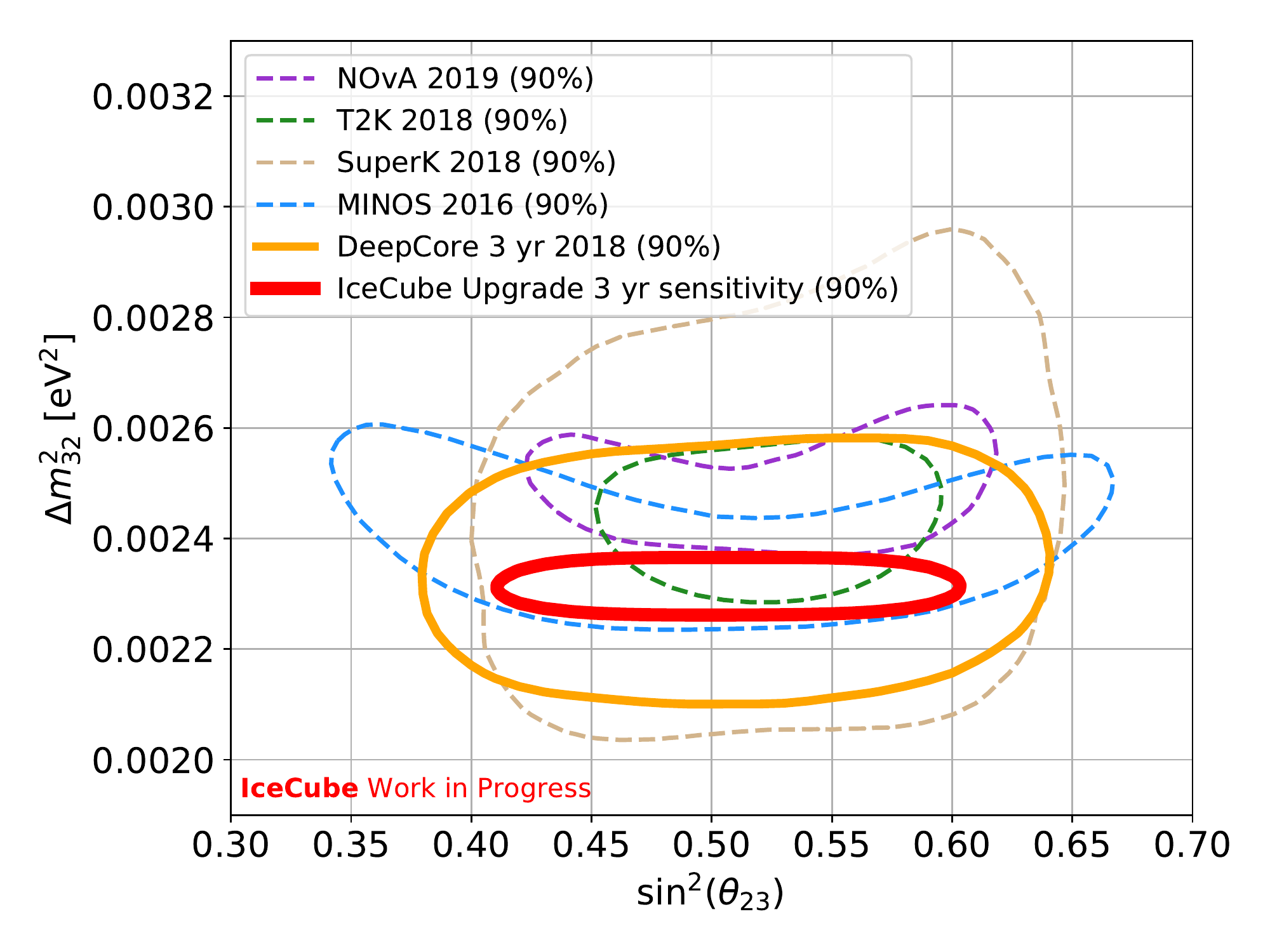}\label{fig:2}}
	\end{minipage}
		\caption{(Left) 68\% sensitivity of the IceCube Upgrade on $\nu_\tau$ normalization value assuming a true value of $1$ with 1 year observation. Also shown are the current best fit values of $\nu_\tau$ normalization from OPERA and Super-Kamiokande. (Right) The predicted performance of the IceCube Upgrade on measurement of sin$^2 \theta_{23}$ and $\Delta m^2_{32}$ assumes 3 years of livetime. Expected 90\% confidence contours in the sin$^2 \theta_{23}$ and $\Delta m^2_{32}$ plane in comparison with the ones of the most sensitive experiments \cite{IC_Oscillations,AtmosNuTau}.}\label{fig:oscillation}
		\end{figure}
According to the standard neutrino oscillation mechanism, atmospheric  disappearance arises primarily from $\nu_\mu\rightarrow\nu_\tau$ oscillations. Here, the $\nu_\mu$ survival probability $P(\nu_\mu\rightarrow\nu_\mu)$ can be approximated as $P(\nu_\mu\rightarrow\nu_\mu) \eqsim1-4|U_{\mu3}|^2 (1-|U_{\mu3}|^2){\rm sin}^2(\frac{\Delta m^2_{32}L}{4E})$, where $U_{\mu3}={\rm sin}\theta_{23}{\rm cos}\theta_{13}$. Also, because atmospheric neutrinos are mostly electron or muon neutrinos at production, tau neutrino appearance is expected in the atmospheric neutrinos from neutrino oscillations. The probability of $\nu_\tau$ appearance is approximated as follows:
$P(\nu_\mu\rightarrow\nu_\tau) \eqsim 4|U_{\mu3}|^2 |U_{\tau3}|^2 {\rm sin}^2(\frac{\Delta m^2_{31}L}{4E})$ where $ 4|U_{\mu3}|^2 |U_{\tau3}|^2 ={\rm sin}^2 2\theta_{23}{\rm cos}^4\theta_{13}$. Neutrino oscillation probabilities depend on the ratio of the path length $L$ to the neutrino energy $E$, allowing the observation of neutrino oscillations as a function of the incident angle (correlated with $L$) and the calculation of their energy. Therefore the reconstruction of the incident neutrino energy and zenith angle is a key ability in the oscillation analysis. For a path length equal to the Earth's diameter, the first oscillation minimum for $\nu_\mu$ and the first oscillation maximum for $\nu_\tau$ are at approximately 25~GeV.

An enhanced photon sensitivity in the Upgrade allows for a more accurate characterization of events during the selection process. 
The upper panels of Fig.~\ref{fig:selection} show the muon and tau neutrino charged current (CC) energy distributions in the Upgrade array compared with those in DeepCore. The figure demonstrates a significant enhancement in the event rates below $\sim$30 GeV. The improvements are observed in the energy region relevant for analyses of neutrino oscillations.  
The ability of IceCube to distinguish $\nu_\mu$ CC interactions, which induces tracks of photon distributions, from the other interactions i.e., $\nu_{e}$ and $\nu_\tau$ CC interactions and neutral current (NC) interactions of $\nu_{e}$, $\nu_{\mu}$ and $\nu_\tau$ neutrinos, which produces only particle shower (cascade) signatures, allows us to measure $\nu_\tau$ contributions in a statistical basis from the simultaneous fitting of track and cascade distributions. The detection efficiency peak of the Upgrade array matches well with the energy range of $\nu_\tau$ oscillation maximum and allows the measurement of a statistically significant number (approximately 3000 events per year) of $\nu_\tau$-induced events. The enhanced sensitivity in oscillation analyses in the Upgrade is the result of both a larger neutrino sample and improved reconstruction performance in these samples as demonstrated in  Fig.~\ref{fig:selection}.

The left panel of Fig.~\ref{fig:oscillation} shows a prediction of the Upgrade sensitivity for $\nu_\tau$ normalization. The Upgrade strings will surpasses the precision of the world's most accurate measurement by a significant amount within approximately one year of operation. Because $\nu_\tau$ appearance and $\nu_\mu$ disappearance probe different elements of the PMNS matrix, a unique and most stringent test of the unitarity of the PMNS matrix can be obtained from the measurements. While the unitarity of a mixing matrix is a necessary condition for the theory with 3 neutrino flavors, experimental tests of unitarity are still considerably poor in the $3\times3$ PMNS matrix. 
Observations of non-unitarity will be the indication of physics beyond the Standard Model. Moreover, the right panel of Fig.~\ref{fig:oscillation} presents the performance of the Upgrade array for the atmospheric neutrino oscillation parameters $\Delta m^2_{23}$ and sin$^2 \theta_{23}$. It shows that the Upgrade will measure those parameters with a precision comparable to that obtained from the current leading accelerator-based experiments, such as the NO$\nu$A, T2K, and MINOS, but with different $L/E$ and systematics. However, we note that future projected sensitivity from these accelerator-based experiments must also be used for fair comparisons.

In the performance study shown here, additional gains from new calibration and reconstruction methods are not included. For instance, an additional improvement could be expected from the calibration of the refrozen ice. The photon scattering from refrozen ice created at the time of the detector deployment is considerably higher. Several devices and strategies have been developed for the calibration of refrozen ice~\cite{POCAM, DEggHoleIce}. 

\subsection{Neutrino Astrophysics}\label{subsec:astrophysics}

In the analysis of neutrinos detected by the IceCube's existing 1 km$^3$ scale photosensor array, the optical properties of the ice are a major source of systematic uncertainty. In the GeV energy region, photon scattering of ice refrozen at the time of detector deployment has a significant contribution to the systematic uncertainties. In the TeV to PeV energy region, because of the increased number of observed scattered photons, the directional reconstruction accuracy of cascades in ice is primary limited by the precise modeling of the bulk layered glacial ice optical properties and by the {\it in situ} response function of the optical sensors~\cite{cascadereco}. In the Upgrade, we will achieve significant improvements on the understanding of the ice and detector response with newly developed calibration devices and optical sensors. Because the event reconstructions are performed based on the arrival time profile and amplitude of Cherenkov photons, the precise measurements of the bulk optical properties of glacial ice as a function of depth is crucial. The depth dependence of the scattering and absorption constants in the glacial ice of IceCube have been studied extensively using 12 LEDs embedded in each of the 5160 IceCube optical modules~\cite{ice}. These parameters are accurately modelled with the theory of Mie scattering with small dust particles in ice. As a secondary effect over the primary depths dependence of the ice properties, a horizontal directional dependence (anisotropy) of the ice parameters was discovered. The observed anisotropy shows difference in scattering and absorption relations compared to the depth-dependent constants. The investigation of this anisotropy led to a description of the refractive scattering of light in the ice crystal boundaries due to the birefringent effect~\cite{ice}. The identification of the possible causes of the newly discovered feature of the South Pole's ice is an important achievement. The model will be further investigated and improved with densely packed photon sensors together with $4\pi$ measurements of light emitted from isotropic light sources and collimated beams, which are capable of sending various wavelengths in any direction. 

 A study that will benefit from the Upgrade calibration is the reconstruction of neutrino-induced cascades. Because of neutrino flavor oscillations over cosmic distances, approximately one 
\begin{wrapfigure}[22]{r}[2pt]{9cm}
\vspace*{-\intextsep} 
	\centering
	\includegraphics[width=0.9\linewidth]{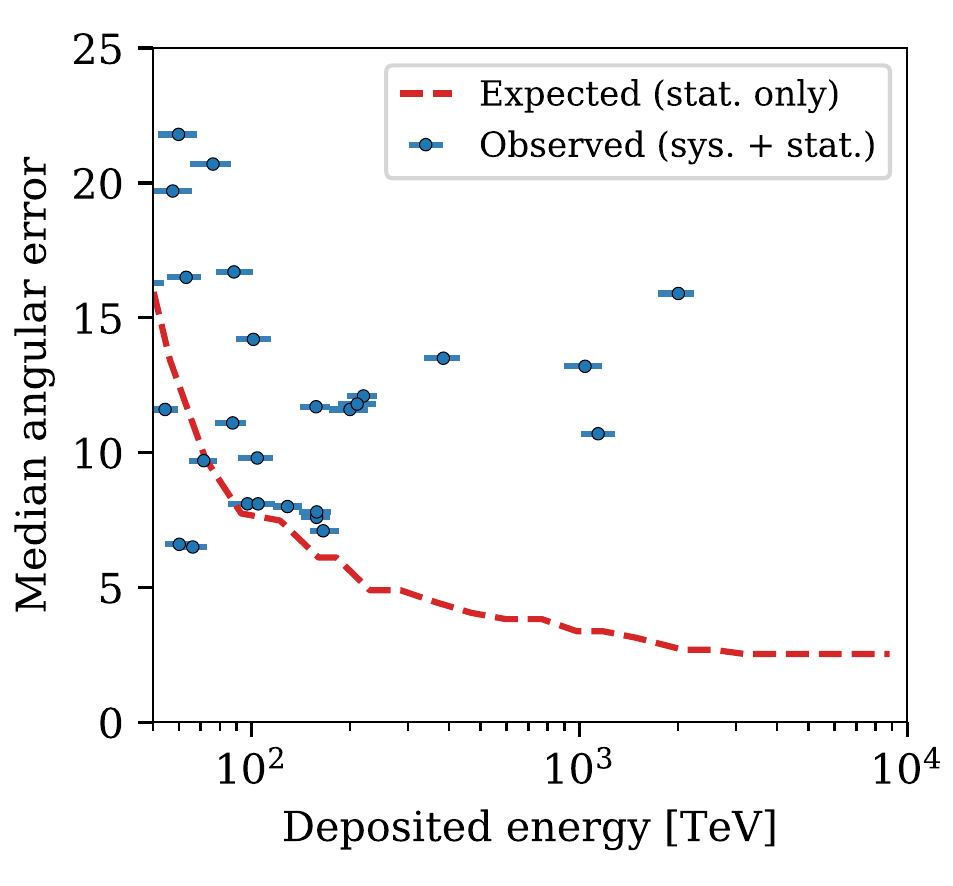}
		\caption{Observed median angular error of fully contained high energy (HESE) cascade directional reconstruction as a function of reconstructed deposited energy. The dashed line indicate the reconstruction performance with a perfect knowledge of the optical properties of ice and detector responses. The deviation of data points from the line indicate the presence of incomplete understandings of ice and detector response to bright light.}
	\label{fig:reconstruction}
\end{wrapfigure}
 third of the cosmic neutrino flux is expected to arrive to Earth as $\nu_{e}$ and another one third as $\nu_\tau$, both of which are detected in IceCube in the form of cascades. Figure~\ref{fig:reconstruction} shows the event-by-event estimates of the angular uncertainty of high-energy neutrino-induced cascades. While cascades without systematic errors can be reconstructed with an uncertainty of 3$^{\circ}$ or less above 1 PeV and 5$^{\circ}$ above 300 TeV, the current reconstruction uncertainty is limited to 10$^{\circ}$ or more in the corresponding energy range, due to the uncertainty on the {\it in situ} detector response and the anisotropy of ice~\cite{cascadereco}. We aim at achieving a cascade angular reconstruction closer to the statistical limit with the planned 
calibration program. The improved cascade directional reconstruction precision will lead to more opportunities for neutrino point source searches using IceCube data collected over the last 10 years. A further improvement on flavor identification is expected for tau neutrinos. 
In high energies, the event-by-event identifications of tau neutrino candidates are possible~\cite{tau}, making use of separation lengths between two cascades, a hadronic cascade in a $\nu_\tau$ CC interaction and an electron or hadronic cascade from the subsequent decay of the tau lepton.
Because tau neutrinos are not expected at the production site of astrophysical neutrinos, their observation provides a unique opportunity to measure neutrino oscillations at cosmological distances and at ultra-high energies. 
An interesting aspect of the flavor ratio is that they are expected to be robust against the flavor composition of the initial astrophysical source and the neutrino oscillation parameters. Deviations from the expectation are unique and robust signatures of new physics. 
While the first $\nu_\tau$ candidates have recently been observed in 7.5 years of IceCube data, tau neutrino identification performance is still limited by ice properties and detector responses. The resultant sensitivity to the flavor composition is insufficient to constrain a hypothesis of new physics. 
An improved precision of the cascade reconstruction as well as tau neutrino flavor identification allows the multi-messenger observations of neutrino-emitting sources and opens up a new way to analyze the flavor dependence of neutrino fluxes.

\subsection{Towards IceCube-Gen2}\label{subsec:gen2}
The observation of a flaring blazar in coincidence with the IceCube real-time alert IC-170922, an extremely high-energy muon neutrino, neutrino astronomy has become a reality. To expand our view of the high-energy Universe through the new window of neutrino astronomy, a next-generation neutrino telescope is highly desired. IceCube-Gen2, currently under design optimization, will consist of approximately 8~km$^3$ of instrumented ice and an array of approximately $1,000$ optical sensors at the South Pole, as an extension of the current IceCube detector. The neutrino point source sensitivity will be enhanced significantly in the IceCube-Gen2. While the Upgrade will deliver compelling science on its own, as described in the previous sections, it will be also useful for the research and development of Gen2, i.e., testing of prototype instrumentation for a full-scale development of the large IceCube-Gen2 array. New deep-ice drilling at the South Pole will enhance the existing IceCube infrastructure for the next generation neutrino telescopes, e.g., {\it in situ} sensitivity study of new sensors, the R\&D study of narrow hole sensor design for significant cost reduction, and verification of large geometrical scale calibration techniques. The Upgrade array will lead to the future next-generation neutrino telescope, IceCube-Gen2.

\bibliographystyle{ICRC}

\end{document}